\documentclass{ws-procs9x6}
\begin{document}

\title{Integral Fluxes, Day-Night, and Spectrum Results from SNO's
  391-Day Salt Phase}

\author{J\"urgen Wendland}

\address{University of British Columbia\\ \emph{for the SNO Collaboration}}

\maketitle

\abstracts{The Sudbury Neutrino Observatory is a $1000\,\mathrm{t}$
  heavy water Cherenkov detector observing neutrinos from the Sun and
  other astrophysical sources.  Measurements of the integral solar
  neutrino fluxes of charged current, neutral current and elastic
  scattering events are reported for 391 days of live data from the
  salt phase of SNO operation. In this phase $2\,\mathrm{t}$ of salt
  were dissolved in the heavy water, which enhanced and differentiated
  the detection of neutral current events.  Day-night asymmetries in
  these fluxes were also determined. The measured electron spectrum
  from the charged-current channel is compatible with the undistorted
  spectrum of the solar $^8\mathrm{B}$ neutrino flux.}

\section{Solar Neutrinos}
Solar neutrinos are produced in nuclear fusion processes in the center
of the Sun, predominantly from the so-called $pp$ cycle.  These
neutrinos have energies up to about $19\,\mathrm{MeV}$. The flux of
solar neutrinos arriving at Earth was measured prior to 2001 by a
variety of experiments sensitive to various ranges in the neutrino
energy spectrum.  These experiments were exclusively or primarily
sensitive to electron neutrinos and included radiochemical
measurements using Cl or Ga and measurements of elastic scattering off
electrons in light water Cherenkov detectors.  The observed solar
neutrino fluxes were two to three times smaller than those predicted
by solar
models\cite{Bahcall:2004fg,Bahcall:2004pz,Turck-Chieze:2004mg}.

This discrepancy may be explained by the oscillation of massive
neutrinos giving rise to flavor change in neutrinos. The formalism for
this process was introduced by Maki, Nakagawa, Sakata, and Pontecorvo
(MNSP)\cite{Maki:1962mu,Gribov:1968kq} and was expanded to include
matter-enhanced oscillations in the Earth or the Sun by Mikheyev,
Smirnov, and Wolfenstein
(MSW)\cite{Mikheev:1986gs,Wolfenstein:1977ue}. In a two neutrino
flavor model the mixing of the two neutrino flavor eigenstates $\nu_e$
and $\nu_\mu$ is described by the mixing angle $\theta$ with respect
to the neutrino mass eigenstates $\nu_1$ and $\nu_2$ and the
difference of the squared neutrino masses $\Delta m^2=m_2^2-m_1^2$.
The survival probability for an electron neutrino in vacuum is then
$P(\nu_e\to\nu_e)=1-\sin^22\theta\sin^2( 1.27\Delta m^2\,L/E)$, where
$L$ is the distance traveled in $\mathrm{km}$ and $E$ the neutrino
energy in $\mathrm{MeV}$. In matter the mixing angle depends on the
electron density $N_e$ of the medium, $\tan^2\theta_M =
\tan\theta/(1-2\sqrt{2}E G_F N_e / (\Delta m^2 \cos2\theta)).$ The
hypothesis of solar neutrino flavor transformation was first directly
verified with measurements by the Sudbury Neutrino Observatory (SNO),
which included a measurement of the total flux of solar $^8\mathrm{B}$
neutrinos via the neutral current reaction in pure heavy
water\cite{Ahmad:2001an,Ahmad:2002jz,Ahmad:2002ka,Ahmed:2003kj}.

\section{The Sudbury Neutrino Observatory}
SNO is a $1000\,\mathrm{t}$ pure heavy water Cherenkov detector
located $2039\,\mathrm{m}$ ($\sim\!6000\,\mathrm{mwe}$) underground in
the Inco Creighton Mine near Sudbury, Ontario,
Canada\cite{Boger:1999bb}.  The heavy water is contained in a
$12\,\mathrm{m}$ diameter acrylic vessel which is surrounded by
$\sim\!7,000\,\mathrm{t}$ of light water. 9,456 photomultiplier tubes
(PMTs) that provide a photocathode coverage of $54\,\%$ are used to
observe neutrino induced Cherenkov events.

SNO detects neutrinos via elastic scattering from electrons (ES,
$\nu_x+e^- \to \nu_x +e^-$), and via the charged current (CC,
$\nu_e+d\hphantom{^-} \to p+p+e^-$ ) and neutral current reactions
(NC, $\nu_x+d\hphantom{^-} \to n+p+\nu_x$) on the deuteron.  The ES
reaction is sensitive to all neutrino flavors, but the electron
neutrino sensitivity is higher than the others by a factor of about
$6.5$. The CC reaction is exclusively sensitive to electron neutrinos,
whereas the NC reaction is equally sensitive to all active neutrino
flavors.

The detection of the neutron is critical for the identification of the
NC reaction. During SNO's first of three phases of operation the
neutron was detected in pure heavy water via photon emission induced
by neutron capture on deuterons. In the second phase, the addition of
$2\,\mathrm{t}$ of salt to the heavy water enhanced the neutron detection
via capture on chlorine and enabled a statistical separation of the NC
and CC events through measurement of the isotropy of events on the
phototubes. For the ongoing third phase, a discrete array of 40 $^3\mathrm{He}$-filled 
proportional counters for individual neutron detection was deployed in
the heavy water.

\section{The SNO Salt Phase}
Neutron capture on $^{35}\mathrm{Cl}$ in the salt increases the
neutron detection efficiency by about a factor three compared to pure
heavy water. In the subsequent de-excitation of $^{36}\mathrm{Cl}$ a
cascade of photons with a total energy of $8.6\,\mathrm{MeV}$ is
released. In comparison to the pure heavy water phase where neutron
capture resulted in a $6.25\,\mathrm{MeV}$ photon, the energy profile
of the radiative capture photopeak is thus moved further above the
analysis threshold. In addition the multi-photon signature of neutron
capture on chlorine is more isotropic than the single-ring Cherenkov
events from the CC and ES interactions. Event light isotropy thus
provides an additional means of distinguishing these event classes.

To obtain the rate of the CC, ES, and NC reactions and the energy
spectrum of the CC events an extended maximum likelihood fit was
applied to 4722 neutrino candidate events from the 391-day salt phase
data set\cite{Aharmim:2005gt}. Probability density functions for each
of the reactions and for an external neutron background were generated
by a detailed Monte Carlo simulation of the detector in terms of the
following parameters: event energy ($T_{eff}$), event direction
($\cos\theta_{\odot}$), volume weighted radius ($\rho=(R/R_{AV})^3$,
$R_{AV}=600.5\,\mathrm{cm}$), and event isotropy ($\beta_{14}$).  The
inclusion of event isotropy allowed a fit to the energy spectrum of
the CC events that was not constrained to the spectral shape of solar
$^8\mathrm{B}$ neutrinos. Systematic uncertainties on the detector
response were measured by comparing data from the Monte Carlo
simulation with data from calibration sources.  By propagating these
uncertainties through the signal extraction process their effects on
the fit parameters were determined.

The fluxes of the CC, ES, and NC channels obtained in this fit are
respectively in units of $10^6\,\mathrm{cm}^{-2}s^{-1}$:
$\phi_{\mathrm{CC}} = 1.68 \pm 0.06\,\text{(stat)}
^{+0.08}_{-0.09}\,\text{(sys)}$, $ \phi_{\mathrm{ES}} = 2.35 \pm
0.22\,\text{(stat)} \pm 0.15\,\text{(sys)}$, and $\phi_{\mathrm{NC}} =
4.94 \pm 0.21\,\text{(stat)} ^{+0.38}_{-0.34}\,\text{(sys)}$.  These
fluxes are the equivalent fluxes of $^8\mathrm{B}$ electron neutrinos
above an energy threshold of zero assuming an undistorted spectral
shape.  The main sources of systematic uncertainties are the isotropy
measurement, the energy scale and bias, the event
reconstruction biases, the neutron detection efficiency (NC flux only)
and the angular resolution (ES flux only). The ratio of the CC and NC
flux is $\phi_{\mathrm{CC}}/\phi_{\mathrm{NC}}=
0.340\pm0.023\text{(stat)}^{+0.029}_{-0.031}\text{(sys)}$, providing
clear evidence that solar electron neutrinos change flavor in transit
to the Earth.

\begin{figure}[t]
  \begin{center}
    \includegraphics[width=0.495\textwidth]{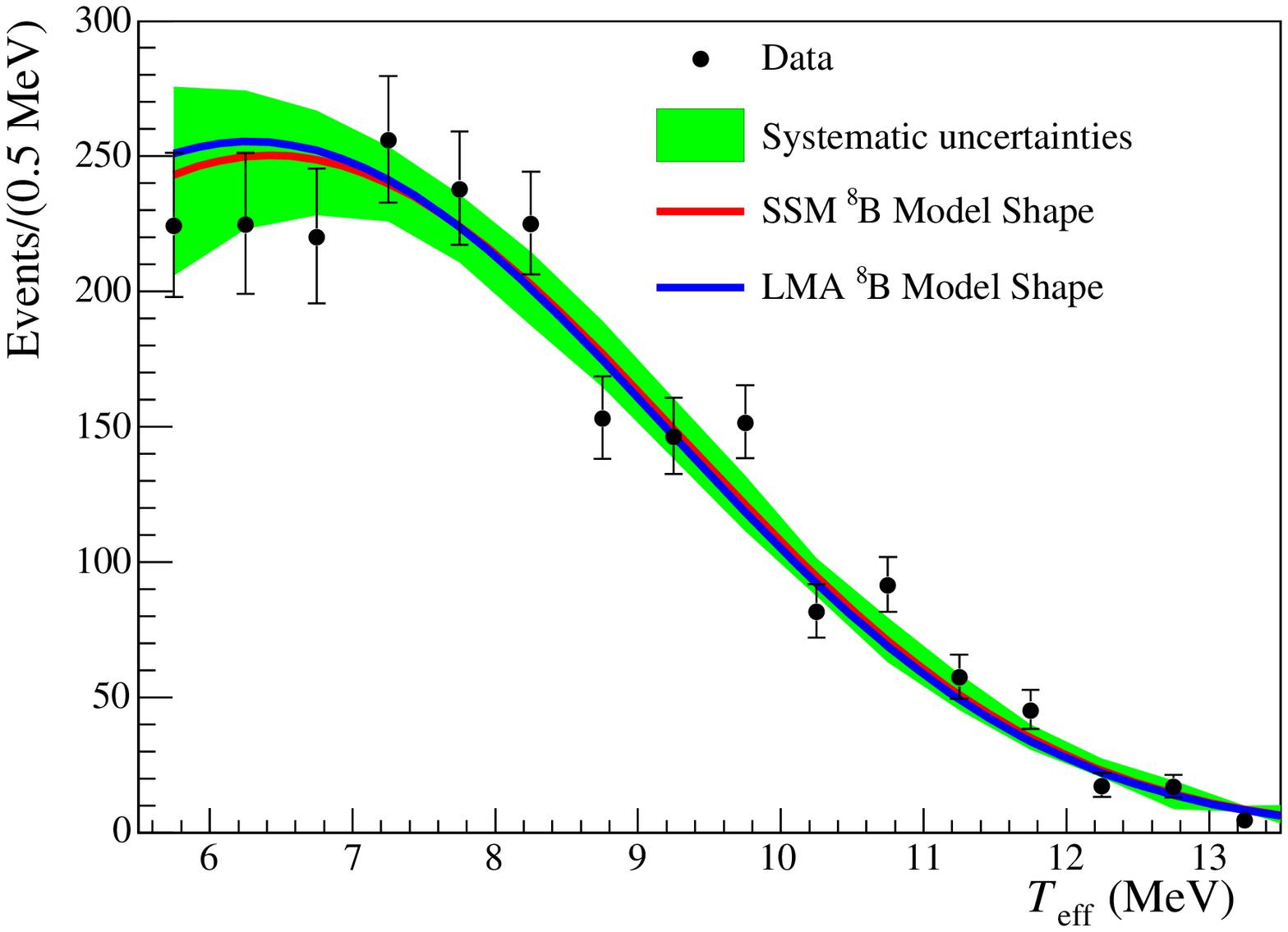}
    \includegraphics[width=.495\textwidth,clip]{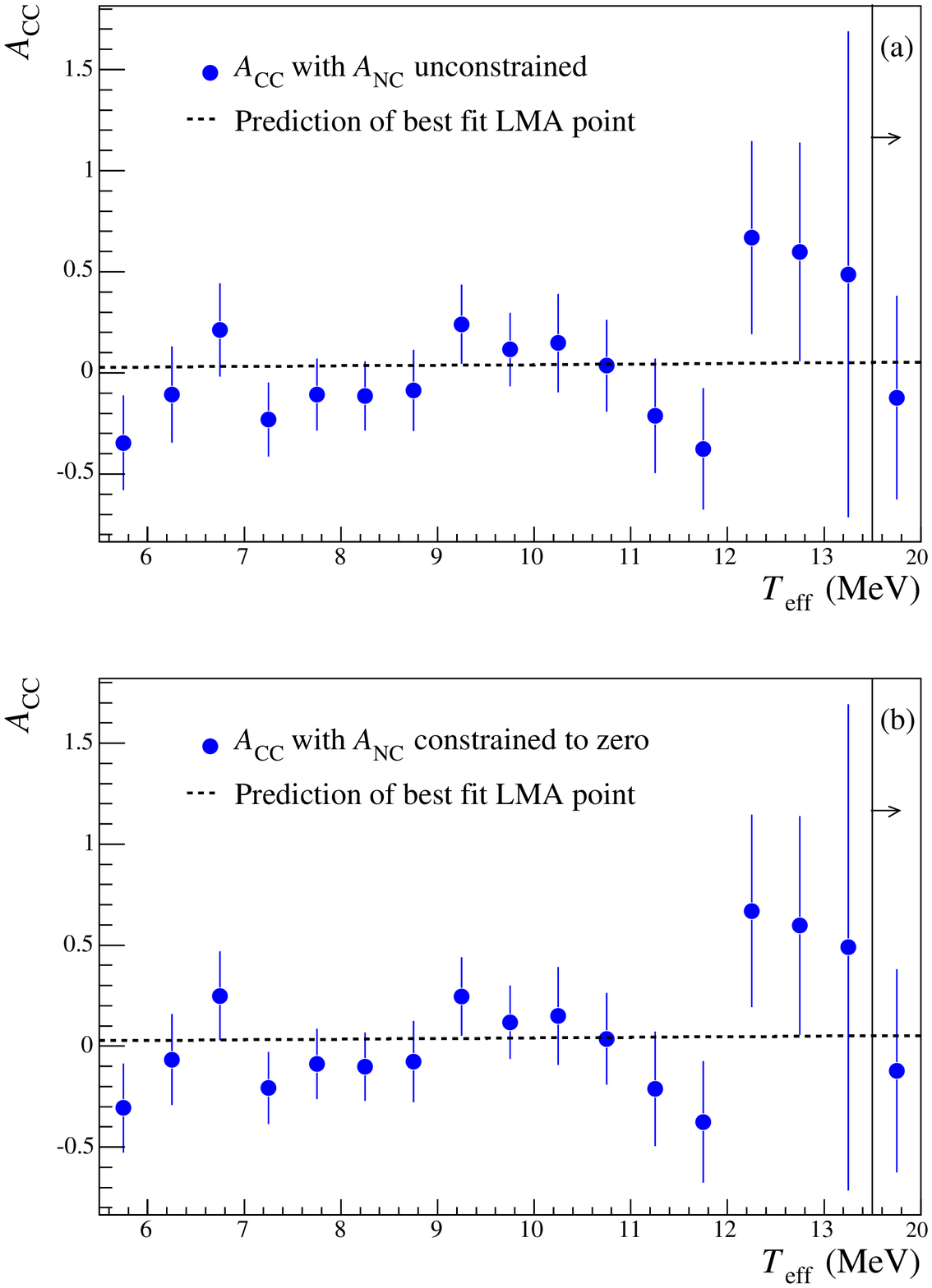}
  \end{center}
  \caption{Left: The energy spectrum of the solar neutrino CC flux. The
    error bars are the statistical uncertainties. The band on the
    undistorted $^8\mathrm{B}$ model shape represent the detector
    systematic uncertainties. Note that the data points are
    statistically and systematically correlated.
    Right: The day-night asymmetry of the charged current solar
    neutrino flux as a function of energy. The errors bars show the
    statistical uncertainties, and the horizontal line shows the
    expectation for the best fit MSW parameters.}
  \label{fig:spectrum}
  \label{fig:dnACC}
\end{figure}

The energy spectrum of the solar CC flux is shown in the left hand
panel of Fig.~\ref{fig:spectrum}. The expected shapes for an
undistorted $^8\mathrm{B}$ solar neutrino flux and for the best fit
MSW model, see below, are also shown.

For certain ranges of mixing parameters the MSW effect predicts a
regeneration of solar electron neutrinos when they pass through the
Earth. The regeneration could be measurable as an asymmetry
$A_{DN}=2\,(\phi_N-\phi_D)/(\phi_N+\phi_D)$ of the solar neutrino flux
$\phi_N$ measured at SNO during the night and the flux measured during
the day $\phi_D$. The energy-unconstrained analysis described above
was carried out separately for the day and night neutrino candidate
events. The resulting asymmetries in the CC, NC, and ES fluxes are
$A_{\mathrm{CC}}=-0.056\pm0.074\text{(stat)}\pm0.053\text{(sys)}$,
$A_{\mathrm{NC}}=0.042\pm0.086\text{(stat)}\pm0.072\text{(sys)}$, and
$A_{\mathrm{ES}}=0.146\pm0.198\text{(stat)}\pm0.033\text{(sys)}$,
respectively, where the systematics largely cancel in the asymmetry
ratio. The day-night asymmetry of the CC flux as a function of
electron energy is shown in the right hand panel of
Fig.~\ref{fig:dnACC}. Within the uncertainties the asymmetry is
compatible with zero and with the best fit MSW solution, discussed in
the following.
\begin{figure}[t]
  \begin{center}
    \includegraphics[width=.5\textwidth,clip]{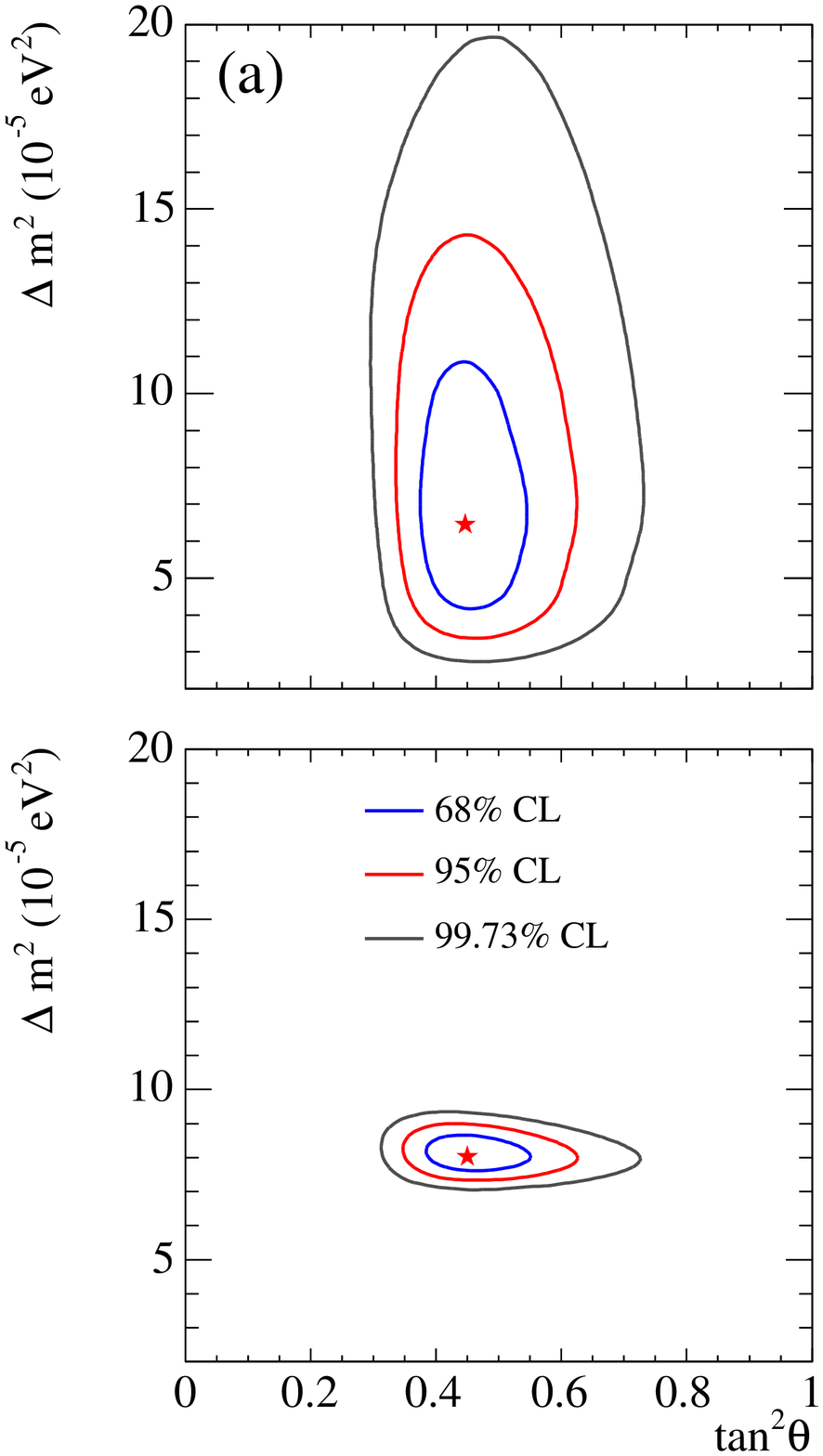}
  \end{center}
  \caption{Neutrino oscillation contours in MSW parameter space for two
    flavor mixing. The best fit point is indicated by the star.}
  \label{fig:msw}
\end{figure}

In an MSW two-parameter fit the presented results were combined with
the global solar neutrino data set from SNO's pure heavy water phase,
the Cl\cite{Cleveland:1998nv} and
Ga\cite{Gavrin:2005ks,Altmann:2000ft} experiments and with the
Super-Kamiokande zenith spectra\cite{Fukuda:2002pe}. Assuming CPT
invariance the rates and spectra of the KamLAND experiment were also
included to further restrict the allowed parameter space.  The result
of this neutrino oscillation analysis is shown in Fig.~\ref{fig:msw}.
The best fit point, $\Delta m^2 =
8.0^{+0.6}_{-0.4}\cdot10^{-5}\,\mathrm{eV}^2$ and
$\theta=33.9\pm^{+2.4}_{-2.2}\,\text{degrees}$, lies in the large
mixing angle (LMA) region. In this fit the SNO data provide strong
constraint of the mixing angle.

\section{Summary}
The SNO collaboration has completed the salt phase data acquisition.
Integral fluxes of the CC, NC, and ES reaction rates were extracted
from the full salt data set. The electron energy spectrum of the solar
electron neutrino flux was observed with the charged current reaction
and the day-night solar neutrino flux asymmetries were measured.  The
spectrum and the day-night asymmetries are consistent with the
no-oscillation hypothesis and with the prediction by the best fit MSW
LMA solution, which was determined in an MSW fit to global solar plus
KamLAND data.


\begin{thebibliography}{0}
\bibitem{Bahcall:2004fg}
  J.~N.~Bahcall and M.~H.~Pinsonneault,
  Phys.\ Rev.\ Lett.\  {\bf 92}, 121301 (2004).

\bibitem{Bahcall:2004pz}
  J.~N.~Bahcall, A.~M.~Serenelli and S.~Basu,
  Astrophys.\ J.\  {\bf 621}, L85 (2005).

\bibitem{Turck-Chieze:2004mg}
  S.~Turck-Chieze {\it et al.},
  Phys.\ Rev.\ Lett.\  {\bf 93}, 211102 (2004).

\bibitem{Maki:1962mu}
  Z.~Maki, M.~Nakagawa and S.~Sakata,
  Prog.\ Theor.\ Phys.\  {\bf 28}, 870 (1962).
\bibitem{Gribov:1968kq}
  V.~N.~Gribov and B.~Pontecorvo,
  Phys.\ Lett.\ B {\bf 28}, 493 (1969).

\bibitem{Mikheev:1986gs}
  S.~P.~Mikheev and A.~Y.~Smirnov,
  Sov.\ J.\ Nucl.\ Phys.\  {\bf 42}, 913 (1985).

\bibitem{Wolfenstein:1977ue}
  L.~Wolfenstein,
  Phys.\ Rev.\ D {\bf 17}, 2369 (1978).


\bibitem{Boger:1999bb}
  J.~Boger {\it et al.}  [SNO],
  Nucl.\ Instrum.\ Meth.\ A {\bf 449}, 172 (2000).

\bibitem{Ahmad:2001an}
  Q.~R.~Ahmad {\it et al.}~[SNO],
  Phys.\ Rev.\ Lett.\ {\bf 87}, 071301 (2001).

\bibitem{Ahmad:2002jz}
  Q.~R.~Ahmad {\it et al.}  [SNO],
  Phys.\ Rev.\ Lett.\  {\bf 89}, 011301 (2002).

\bibitem{Ahmad:2002ka}
  Q.~R.~Ahmad {\it et al.}  [SNO],
  Phys.\ Rev.\ Lett.\  {\bf 89}, 011302 (2002).

\bibitem{Ahmed:2003kj}
  S.~N.~Ahmed {\it et al.}  [SNO],
  Phys.\ Rev.\ Lett.\  {\bf 92}, 181301 (2004).

\bibitem{Aharmim:2005gt}
  B.~Aharmim {\it et al.}~[SNO],
  arXiv:nucl-ex/0502021.

\bibitem{Cleveland:1998nv}
  B.~T.~Cleveland {\it et al.},
  Astrophys.\ J.\  {\bf 496}, 505 (1998).

\bibitem{Gavrin:2005ks}
  V.~N.~Gavrin~[SAGE],
  Nucl.\ Phys.\ Proc.\ Suppl.\  {\bf 138}, 87 (2005).

\bibitem{Altmann:2000ft}
  M.~Altmann {\it et al.}~[GNO],
  Phys.\ Lett.\ B {\bf 490}, 16 (2000).

\bibitem{Fukuda:2002pe}
  S.~Fukuda {\it et al.}~[Super-Kamiokande],
  Phys.\ Lett.\ B {\bf 539}, 179 (2002).

\end{thebibliography}
\end{document}